\documentstyle[preprint,aps,epsf]{revtex}
\newcommand {\bea}{\begin{eqnarray}}
\newcommand {\eea}{\end{eqnarray}}
\newcommand {\be}{\begin{equation}}
\newcommand {\ee}{\end{equation}}

\begin{document}

%\draft
\preprint{SUNY-NTG-02-03}

\title{Instanton Effects in QCD at High Baryon Density}

\author{T.~Sch\"afer$^{1,2,3}$}

\address{
$^1$Department of Physics, Duke University, Durham, NC 27708\\
$^2$Department of Physics, SUNY Stony Brook, Stony Brook, NY 11794\\ 
$^3$Riken-BNL Research Center, Brookhaven National Laboratory, 
Upton, NY 11973}

\maketitle
\begin{abstract}

  We study instanton effects in QCD at very high baryon density.
In this regime instantons are suppressed by a large power of 
$(\Lambda_{QCD}/\mu)$, where $\Lambda_{QCD}$ is the QCD scale
parameter and $\mu$ is the baryon chemical potential. Instantons
are nevertheless important because they contribute to several
physical observables that vanish to all orders in perturbative
QCD. We study, in particular, the chiral condensate and its 
contribution $m_{GB}^2\sim m\langle\bar{\psi}\psi\rangle$ to 
the masses of Goldstone bosons in the CFL phase of QCD with 
$N_f=3$ flavors. We find that at densities $\rho\sim (5-10)
\rho_0$, where $\rho_0$ is the density of nuclear matter, the 
result is dominated by large instantons and subject to considerable 
uncertainties. We suggest that these uncertainties can be addressed 
using lattice calculations of the instanton density and the pseudoscalar 
diquark mass in QCD with two colors. We study the topological 
susceptibility and Witten-Veneziano type mass relations in both 
$N_c=2$ and $N_c=3$ QCD. 

\end{abstract}

\newpage

%%%%%%%%%%%%%%%%%%%%%%%%%%%%%%%%%%%%%%%%%%%%%%%%%%%%%%%%%%%%%%%%%%%%%%%%%
\section{Introduction}
\label{sec_intro}
%%%%%%%%%%%%%%%%%%%%%%%%%%%%%%%%%%%%%%%%%%%%%%%%%%%%%%%%%%%%%%%%%%%%%%%%%

  Strange quark matter at very high baryon density but low
temperature is believed to be in the color-flavor-locked (CFL) phase 
\cite{Alford:1999mk,Schafer:1999fe,Evans:2000at}. The CFL phase 
is characterized by diquark condensates which break both the 
global $SU(3)_L\times SU(3)_R$ flavor symmetry and the local $SU(3)$ 
color symmetry but preserve a global $SU(3)_V$. In addition to 
that, diquark condensation spontaneously breaks the exact 
$U(1)_V$ and approximate $U(1)_A$ symmetry of QCD. The low
energy behavior of the CFL phase is governed by the corresponding
Goldstone bosons, an octet of pseudoscalar mesons associated
with chiral symmetry breaking and two singlets associated
with the breaking of $U(1)_{V,A}$. 

 The CFL phase has many similarities with $N_f=3$ QCD at low
density \cite{Alford:1999mk,Schafer:1999ef}, but the mechanism of 
chiral symmetry breaking is quite different. In particular, if 
the baryon density is very large, the dominant order parameter 
for chiral symmetry breaking is not the usual quark-anti-quark 
condensate $\langle\bar{\psi}\psi\rangle$, but a four-fermion operator 
$\langle (\bar{\psi}\psi)^2\rangle \sim \langle \psi\psi\rangle^2$. 
As a consequence, the masses squared of the pseudoscalar Goldstone 
bosons are quadratic, not linear, in the quark masses. The coefficient 
of proportionality can be computed in perturbative QCD
\cite{Son:1999cm,Rho:2000xf,Hong:2000ei,Manuel:2000wm,Rho:2000ww,Beane:2000ms,Hong:2000ng,Schafer:2001za}.

 The CFL diquark condensate leaves a discrete axial $(Z_2)_A$ symmetry
unbroken. Since the quark-anti-quark condensate is odd under this 
symmetry, $\langle\bar{\psi}\psi\rangle$ remains zero in weak
coupling perturbation theory. Non-perturbative effects, instantons,
break the $(Z_2)_A$ symmetry and lead to a non-zero expectation
value $\langle\bar{\psi}\psi\rangle$. If the density is large 
instantons are strongly suppressed and $\langle\bar{\psi}\psi\rangle$ 
is small. As the density decreases instantons become more important 
and the chiral condensate grows. Our main objective in the present 
work is to compute the linear term in the mass relation for the 
Goldstone bosons. The relative size of the linear and quadratic terms 
in the mass relation provides an estimate of the baryon density
at which the crossover from the asymptotic regime, in which chiral 
symmetry breaking is dominated by the diquark condensate, to the low 
density regime, in which chiral symmetry breaking is governed by the 
quark-anti-quark condensate, occurs. 

 The size of the linear term in the Goldstone boson mass
relation also has important consequences for the phase structure
of CFL matter in the presence of a non-zero strange quark mass
and lepton chemical potentials \cite{Alford:1999pa,Schafer:1999pb,Alford:2001ze,Schafer:2000ew,Bedaque:2001je,Alford:2001zr,Kaplan:2001qk}.
Bedaque and Sch\"afer argued that for physical values of the quark 
masses and the baryon chemical potential CFL quark matter is likely 
to be kaon condensed \cite{Bedaque:2001je}. This conclusion was 
based on an analysis of the effective potential for the chiral 
order parameter. The low energy constants that appear in the 
effective potential were obtained in perturbation theory, but 
the conclusions are unchanged if the coefficients are estimated 
using dimensional analysis. This suggests that the results of the 
perturbative calculation are qualitatively correct even in the 
regime of strong coupling. However, the assumption that instanton 
effects are small for the densities of interest is crucial. 
 
 The paper is organized as follows. In section \ref{sec_inst} we 
compute the instanton contribution to Goldstone boson masses in 
the CFL phase. In section \ref{sec_size} we study corrections to
this result due to the finite size of instantons. In section 
\ref{sec_nc2} we compute the pseudoscalar diquark mass in QCD 
with two colors. This observable can be studied with present 
lattice techniques. We also comment on Witten-Veneziano relations 
in QCD at high baryon density. Our results extend previous work 
in \cite{Rapp:1999qa,Schafer:1999fe,Manuel:2000wm,Son:2001jm}.

%%%%%%%%%%%%%%%%%%%%%%%%%%%%%%%%%%%%%%%%%%%%%%%%%%%%%%%%%%%%%%%%%%%%%%%%%
\section{Instanton Contribution to CFL chiral theory}
\label{sec_inst}
%%%%%%%%%%%%%%%%%%%%%%%%%%%%%%%%%%%%%%%%%%%%%%%%%%%%%%%%%%%%%%%%%%%%%%%%%

\begin{figure}
\begin{center}
\leavevmode
\vspace*{0cm}
\epsfxsize=12cm
\epsffile{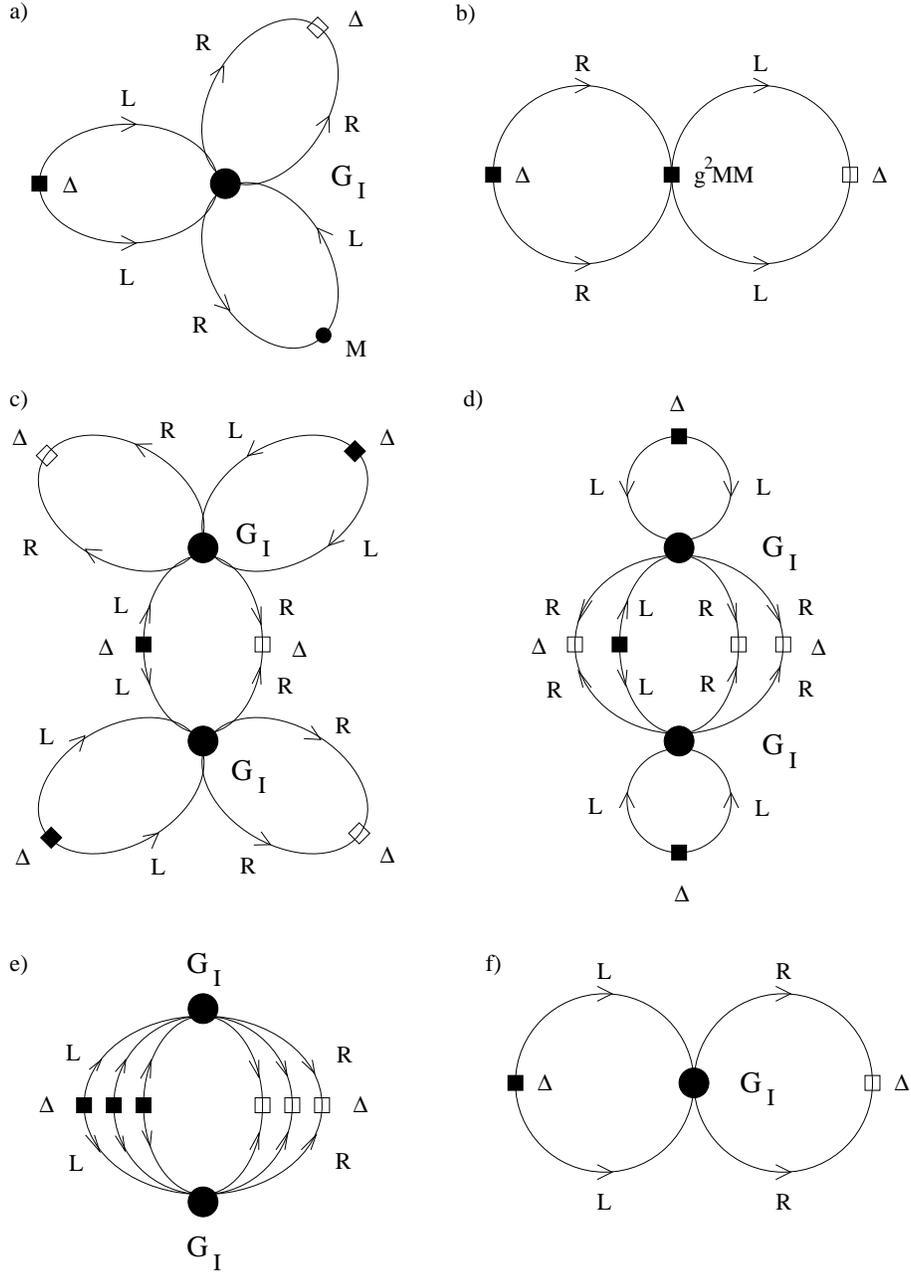}
\vspace*{0cm}
\end{center}  
\caption{\label{fig_mvac}
Figure a) shows the $O(M)$ instanton contribution to the vacuum 
energy in the CFL phase.  The six-fermion vertex is the effective 
't Hooft vertex in the field of an instanton. Figure b) shows the 
$O(M^2)$ perturbative contribution to the vacuum energy. The 
four-fermion vertex corresponds to hard ($p\sim p_F$) electric 
gluon exchange. Figures c),d) and e) show the two-instanton
contribution to the vacuum energy in the chiral limit.
Figure f) shows the instanton contribution to the 
vacuum energy in QCD with two colors and two flavors.}
\end{figure}

 In this section we consider the CFL phase of high density quark
matter. For excitation energies smaller than the gap the only relevant 
degrees of freedom are the Goldstone modes associated with the breaking 
of chiral symmetry and baryon number. The interaction of the Goldstone 
modes is described by the effective Lagrangian \cite{Casalbuoni:1999wu}
\bea
\label{l_cheft}
{\cal L}_{eff} &=& \frac{f_\pi^2}{4} {\rm Tr}\left[
 \nabla_0\Sigma\nabla_0\Sigma^\dagger - v_\pi^2
 \partial_i\Sigma\partial_i\Sigma^\dagger \right]
 + \frac{3f^2}{4} \left[ 
 \partial_0 V\partial_0 V^*  - v_{\eta'}^2 
 \partial_i V\partial_i V^* \right]  \\
 & & \hspace{-1cm}\mbox{}
     +\Big[ A {\rm Tr}(M\Sigma^\dagger)V^* + h.c. \Big] \nonumber \\
 & & \hspace{-1cm}\mbox{} 
     +\Big[ B_1{\rm Tr}(M\Sigma^\dagger)
                        {\rm Tr} (M\Sigma^\dagger)V 
     + B_2{\rm Tr}(M\Sigma^\dagger M\Sigma^\dagger)V  
     + B_3{\rm Tr}(M\Sigma^\dagger){\rm Tr} (M^\dagger\Sigma)
         + h.c. \Big]+\ldots . \nonumber
\eea
Here $\Sigma=\exp(i\phi^a\lambda^a/f_\pi)$ is the chiral field,
$f_\pi$ and $f$ are the octet and singlet decay constants, $M$ 
is a complex mass matrix. The chiral field and the mass matrix 
transform as $\Sigma\to L\Sigma R^\dagger$ and  $M\to LMR^\dagger$ 
under chiral transformations $(L,R)\in SU(3)_L\times SU(3)_R$. 
The axial $U(1)_A$ field is $V=\exp(i\phi)=\exp(2i\eta'/(\sqrt{6}f))$. 
As explained in \cite{Bedaque:2001je} the covariant derivative 
$\nabla_0\Sigma$ contains the effective chemical potentials 
$X_L=(M M^\dagger)/(2\mu)$ and $X_R=(M^\dagger M)/(2\mu)$, 
\be
\label{cov}
\nabla_0\Sigma = \partial_0 \Sigma 
 + i \left(\frac{M M^\dagger}{2\mu}\right)\Sigma
 - i \Sigma\left(\frac{ M^\dagger M}{2\mu}\right) .
\ee
The coefficients $B_i$ of the quadratic mass terms were computed in 
\cite{Son:1999cm,Rho:2000xf,Hong:2000ei,Manuel:2000wm,Rho:2000ww,Beane:2000ms,Hong:2000ng,Schafer:2001za}
by matching the mass dependence of the vacuum energy computed in 
perturbative QCD and the CFL chiral theory. The result is 
\cite{Son:1999cm,Schafer:2001za}
\be 
\label{O(m2)}
 B_1= -B_2 = \frac{3\Delta^2}{4\pi^2}, 
\hspace{1cm} B_3 = 0,
\ee
where $\Delta$ is the gap in the quasi-particle spectrum. 
We shall determine the coefficient $A$ by computing the 
shift in the vacuum energy linear in the quark mass. This
shift is due to the first order instanton contribution to 
the vacuum energy in the background of a perturbatively generated 
diquark condensate, see the diagram shown in Fig.~\ref{fig_mvac}a.
In QCD with three flavors, the instanton induced interaction
between quarks is given by \cite{'tHooft:up,Shifman:uw,Schafer:1996wv}
\bea
\label{l_nf3}
{\cal L} &=& \int n(\rho,\mu)d\rho\, 
 \frac{(2\pi\rho)^6\rho^3}{6N_c(N_c^2-1)}
 \epsilon_{f_1f_2f_3}\epsilon_{g_1g_2g_3}
 \left( \frac{2N_c+1}{2N_c+4}
  (\bar\psi_{R,f_1} \psi_{L,g_1})
  (\bar\psi_{R,f_2} \psi_{L,g_2})
  (\bar\psi_{R,f_3} \psi_{L,g_3}) \right. \\
& & \mbox{}\left. - \frac{3}{8(N_c+2)}
  (\bar\psi_{R,f_1} \psi_{L,g_1})
  (\bar\psi_{R,f_2} \sigma_{\mu\nu} \psi_{L,g_2})
  (\bar\psi_{R,f_3} \sigma_{\mu\nu} \psi_{L,g_3})
  + ( L \leftrightarrow R ) \right) \nonumber .
\eea
Here, $\rho$ is the instanton size, $\mu$ is the quark chemical
potential, $f_i,g_i$ are flavor indices and $\sigma_{\mu\nu}
=\frac{i}{2}[\gamma_\mu,\gamma_\nu]$. The instanton size
distribution $n(\rho,\mu)$ is given by
\bea
\label{G_I}
  n(\rho,\mu) &=& C_{N} \ \left(\frac{8\pi^2}{g^2}\right)^{2N_c} 
 \rho^{-5}\exp\left[-\frac{8\pi^2}{g(\rho)^2}\right]
 \exp\left[-N_f\rho^2\mu^2\right],\\
 && C_{N} \;=\; \frac{0.466\exp(-1.679N_c)1.34^{N_f}}
    {(N_c-1)!(N_c-2)!}\, ,\\
 && \frac{8\pi^2}{g^2(\rho)} \;=\; 
    -b\log(\rho\Lambda), \hspace{1cm} 
    b = \frac{11}{3}N_c-\frac{2}{3}N_f \, . 
\eea
At zero density, the $\rho$ integral in equ.~(\ref{G_I}) is
divergent at large $\rho$. This is the well-known infrared 
problem of the semi-classical approximation in QCD. At 
large chemical potential, however, large instantons 
are suppressed and the typical instanton size is $\rho
\sim \mu^{-1} \ll \Lambda^{-1}$. To linear order in the 
quark mass one of the three zero modes is lifted. We 
obtain 
\bea
\label{l_nf2}
{\cal L} &=& \int n(\rho,\mu)d\rho\, 
   \frac{2(2\pi\rho)^4\rho^3}{4(N_c^2-1)}
 \epsilon_{f_1f_2f_3}\epsilon_{g_1g_2g_3}M_{f_3 g_3}
 \left( \frac{2N_c-1}{2N_c}
  (\bar\psi_{R,f_1} \psi_{L,g_1})
  (\bar\psi_{R,f_2} \psi_{L,g_2}) \right. \\
& & \hspace{1cm}\mbox{}\left. - \frac{1}{8N_c}
  (\bar\psi_{R,f_1} \sigma_{\mu\nu} \psi_{L,g_1})
  (\bar\psi_{R,f_2} \sigma_{\mu\nu} \psi_{L,g_2})
  + (M\leftrightarrow M^\dagger, 
     L \leftrightarrow R ) \right) \nonumber ,
\eea
We can now compute the expectation value of equ.~(\ref{l_nf2}) in 
the CFL ground state \cite{Schafer:1999fe,Rapp:1999qa,Son:2001jm}. 
Using the perturbative result for the diquark condensate in the 
CFL phase,
\bea
\label{qq_cond}
\langle \psi^a_{L,f} C\psi^b_{L,g}\rangle 
  &=& -\langle \psi^a_{R,f} C\psi^b_{R,g}\rangle = 
 \left( \delta^a_f\delta^b_g-\delta^a_g\delta^b_f \right) \Phi,\\
 & & \Phi \,=\,   \frac{3\sqrt{2}\pi}{g} \Delta 
     \left(\frac{\mu^2}{2\pi^2}\right), \nonumber
\eea
we find the instanton contribution to the vacuum energy density
\bea
\label{E_I}
{\cal E} &=& -\int n(\rho,\mu) d\rho\,
 \frac{16}{3}(\pi\rho)^4\rho^3 
 \left[\frac{3\sqrt{2}\pi}{g}\Delta
     \left(\frac{\mu^2}{2\pi^2}\right)\right]^2
  {\rm Tr}\left[M+M^\dagger\right] . 
\eea
We note that for $M={\rm diag}(m_u,m_d,m_s)$ the instanton
contribution to the vacuum energy is indeed negative. Since the
effective interaction involves both left and right-handed 
fermions the relative phase between the left and right-handed
condensate in equ.~(\ref{qq_cond}) is important. Instantons 
favor the state with $\langle\psi_L\psi_L\rangle = -\langle
\psi_R\psi_R\rangle$ which is the parity even ground state. 
Equation (\ref{E_I}) for the vacuum energy can be matched against 
the effective lagrangian equ.~(\ref{l_cheft}). We find
\be 
\label{A}
 A = C_N\frac{8\pi^4}{3}\frac{\Gamma(6)}{3^6}
 \left[\frac{3\sqrt{2}\pi}{g}\Delta
     \left(\frac{\mu^2}{2\pi^2}\right)\right]^2
  \left(\frac{8\pi^2}{g^2}\right)^{6}
  \left(\frac{\Lambda}{\mu}\right)^{12}\Lambda^{-3} ,
\ee
where we have performed the integral over the instanton
size $\rho$ using the one-loop beta function. We note that
$A$ is related to the quark-anti-quark condensate, $\langle
\bar{\psi}\psi\rangle =-2A$. Equation (\ref{A}) agrees, up
to a numerical factor and a power of $g$, with the estimate 
presented in \cite{Schafer:1999fe,Manuel:2000wm}. We can also 
determine the masses of Goldstone bosons. We take into account 
the instanton contribution equ.~(\ref{A}), the $O(M^2)$ term 
given in equ.~(\ref{O(m2)}) and the $O(M^4)$ term given in 
equ.~(\ref{cov}). To this order, the masses of the charged 
Goldstone bosons are given by
\bea 
\label{mgb}
 m_{\pi^\pm} &=&  \mp\frac{m_d^2-m_u^2}{2\mu} +
         \left[\frac{2A}{f_\pi^2}(m_u+m_d)+
               \frac{4B}{f_\pi^2}(m_u+m_d)m_s\right]^{1/2},\nonumber \\
 m_{K_\pm}   &=&  \mp \frac{m_s^2-m_u^2}{2\mu} + 
         \left[\frac{2A}{f_\pi^2}(m_u+m_s)+
               \frac{4B}{f_\pi^2}m_d (m_u+m_s)\right]^{1/2}, \\
 m_{K^0,\bar{K}^0} &=&  \mp \frac{m_s^2-m_d^2}{2\mu} + 
         \left[\frac{2A}{f_\pi^2}(m_d+m_s)+
               \frac{4B}{f_\pi^2}m_u (m_d+m_s)\right]^{1/2}.\nonumber
\eea
In the flavor symmetric limit $m_u=m_d=m_s\equiv m$ the one-instanton 
contribution to the mass of the $\eta'$ is 
\be
\label{m_eta}
 m_{\eta'}^2 =  \frac{4A}{f^2} m.
\ee
If flavor symmetry is broken the $\eta'$ mixes with the $\eta$
and $\pi^0$ \cite{Manuel:2000wm,Beane:2000ms,Schafer:2001za}.
We note that at the one-instanton level, the mass of the $\eta'$ 
vanishes in the chiral limit. The $\eta'$-mass in the chiral
limit arises from the two-instanton diagrams shown in 
Figs.~\ref{fig_mvac}c)-\ref{fig_mvac}e). These diagrams are 
hard to evaluate and we will not pursue this problem here. 
However, even without a calculation we can determine the 
dependence of the instanton generated potential on the QCD 
theta angle $\theta$ and the $U(1)_A$ phase $\phi$ of the 
chiral field. Again restricting ourselves to the case of 
exact flavor symmetry we find
\be
\label{v_inst}
 V = -6mA\cos(\theta+\phi)-12m^2 B\cos(\phi) 
      -2C\cos(2\theta+3\phi),
\ee
where $A$ is the coefficient of the one-instanton contribution
given in equ.~(\ref{A}), $B=B_1=B_2$ is the coefficient of the 
$O(M^2)$ term given in equ.~(\ref{O(m2)}), and $C$ is the coefficient 
of the two-instanton term. The potential equ.~(\ref{v_inst}) 
determines the topological susceptibility in the CFL phase. 
In the limit of very small quark masses we find
\be
\chi_{top} = \frac{2mA}{3} = 
 -\frac{m\langle\bar{\psi}\psi\rangle}{3}.
\ee
This result agrees with prediction of anomalous Ward identities 
at zero density. If we take the chemical potential to infinity 
while keeping the quark mass fixed then $m^2B\gg mA \gg C$ and 
the topological susceptibility is given by $\chi_{top}=6mA=-3m
\langle \bar{\psi}\psi\rangle$.

\begin{figure}
\begin{center}
\leavevmode
\vspace*{0cm}
\epsfxsize=12cm
\epsffile{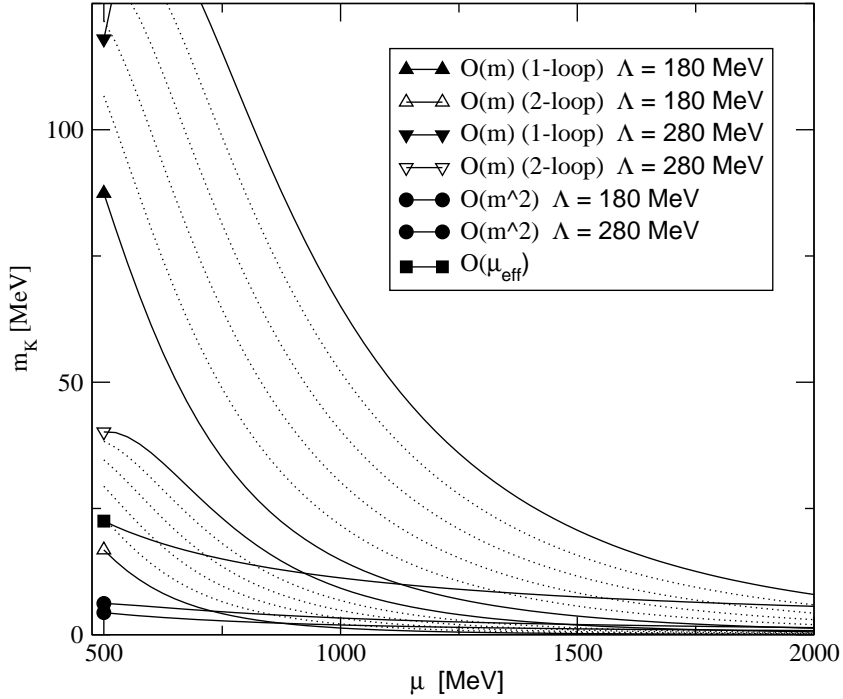}
\end{center}  
\caption{\label{fig_mgb_1}
Different contributions to the $K^0$ mass at finite baryon density. 
The two sets of dotted lines bounded by solid lines marked with triangles 
show the instanton contribution to the kaon mass computed for different 
values of the $\Lambda$ parameter in the range $\Lambda=(180-280)$ MeV. 
The curves marked with solid (open) triangles are computed with the 
one-loop (two-loop) instanton distribution. The solid lines marked with
circles show the perturbative $O(m^2)$ contribution for two 
different values of the scale parameter. The solid line marked
by a square shows the absolute magnitude of the effective chemical 
potential $O(\mu_{eff}^2)=O(m^4)$ contribution. }
\end{figure}

\begin{figure}
\begin{center}
\leavevmode
\vspace*{-1cm}
\epsfxsize=12cm
\epsffile{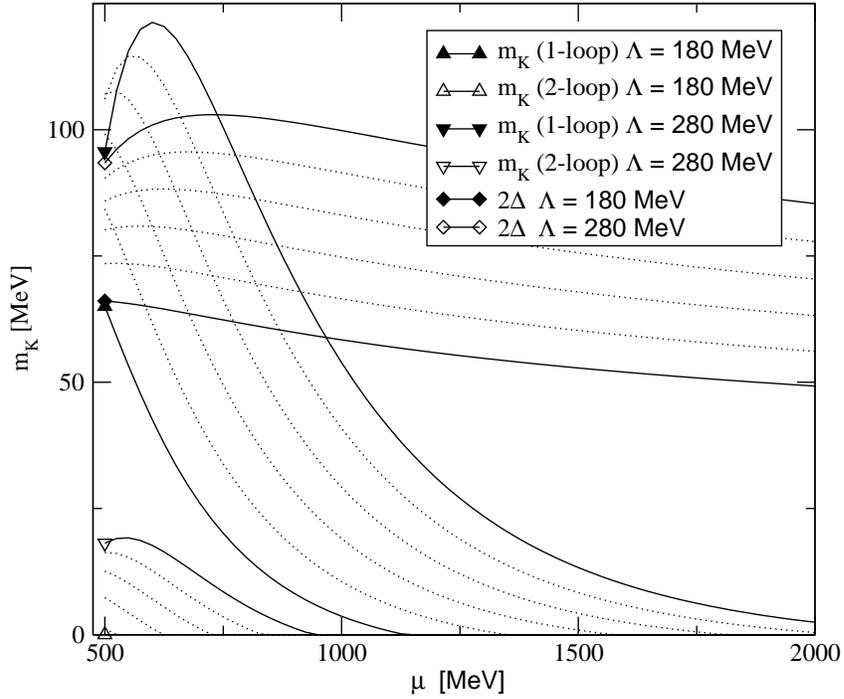}
\end{center}  
\caption{\label{fig_mgb_2}
This figure shows the sum of the different contributions to 
the $K^0$ mass shown in Fig.~\ref{fig_mgb_1}. For comparison 
we also show the gap $2\Delta$ in the excitation spectrum for 
different values of the scale parameter in the range $\Lambda=
(180-280)$ MeV.}
\end{figure}

 In the following, we will study the mass of the $K^0$ in more 
detail. The mass of the $K^0$ receives the largest instanton 
contribution and is of special interest in connection with kaon 
condensation. Different contributions to the $K^0$-mass are show 
in Fig.~\ref{fig_mgb_1}. The $O(m)$ and $O(m^2)$ contributions
are computed from equ.~(\ref{mgb}) taking into account only 
the terms proportional to $A$ and $B$, respectively. The $O(m^4)$
term $(m_s^2-m_d^2)/(2\mu)$ increases the mass of the $\bar{K}^0$,
but decreases the mass of the $K_0$. The complete result for
the $K^0$ mass is show in Fig.~\ref{fig_mgb_2}. If the $O(m^4)$
term is bigger than the $O(m,m^2)$ contribution then $K^0$ 
condensation takes place and the mass of the $K^0$ vanishes. 
We have used $m_u=4$ MeV, $m_d=7$ MeV, $m_s=150$ MeV and the 
perturbative result for the gap
\cite{Son:1999uk,Schafer:1999jg,Hong:2000fh,Pisarski:2000tv}
\be 
\label{gap}
 \Delta = 512\pi^4 2^{-1/3} (2/3)^{-5/2} b_0'\mu g^{-5}
 \exp\left(-\frac{3\pi^2}{\sqrt{2}g}\right),
\ee 
where $b_0'$ is a constant which is determined by non-Fermi liquid
effects \cite{Brown:1999aq,Wang:2001aq} that are not included in 
our calculation. To leading order in perturbation theory the 
pion decay constant is given by
\cite{Son:1999cm,Beane:2000ms,Zarembo:2000pj,Miransky:2001bd} 
\be
f_\pi^2 \,=\, \frac{21-8\log(2)}{18} 
  \left(\frac{\mu^2}{2\pi^2} \right).
\ee
Qualitatively, the instanton contribution to the kaon mass
scales as
\be
 \left. m_{K}\right|_{inst} \sim
  \left(\frac{m_s}{\Lambda}\right)^{1/2}
  \left(\frac{\Delta}{\Lambda}\right)
  \left(\frac{\Lambda}{\mu}\right)^5
  \left[\log\left(\frac{\mu}{\Lambda}\right)\right]^{7/2}
  \Lambda.
\ee
This shows that the result is strongly suppressed as
$\mu\to\infty$. We also note, however, that the result
is quite sensitive to the value of the scale parameter
$\Lambda$. In practice, the power dependence on the 
scale parameter is canceled to some degree by the 
logarithmic dependence. This can be seen from the results shown 
in in Fig.~\ref{fig_mgb_1}. At $\mu=500$ MeV, the instanton
contribution to $m_{K^0}$ calculated from equ.~(\ref{A})
varies between 85 MeV and 120 MeV if the scale parameter
is varied between 180 MeV and 280 MeV.

 The dependence of $m_{K^0}$ on the scale parameter provides
a naive estimate of the importance of higher order corrections. 
In the present case this is probably an underestimate. This
can be seen by studying the role of higher order corrections
in the instanton size distribution. At two-loop order we have
\bea
\label{2loop}
  n_{II}(\rho,\mu) &=& C_{N} \ \left(\beta_I(\rho)\right)^{2N_c} 
 \rho^{-5}\exp\left[-\beta_{II}(\rho)-N_f\rho^2\mu^2\right],\\
 && \beta_I(\rho) \,\,\;=\;
    -b\log(\rho\Lambda), \hspace{3.1cm} 
    b = \frac{11}{3}N_c-\frac{2}{3}N_f \, , \\
 && \beta_{II}(\rho) \;=\;
    \beta_I(\rho) + \frac{b'}{2b}
    \log\left(\frac{2\beta_I(\rho)}{b}\right), \hspace{0.5cm} 
    b' = \frac{34}{3}N_c^2-\frac{13}{3}N_fN_c +\frac{N_f}{N_c} \, . 
\eea
The instanton contribution to the kaon mass calculated with the
two-loop instanton distribution is also shown in Fig.~\ref{fig_mgb_1}.
We observe that the results are significantly smaller as compared
to the leading order estimate. We find $m_{K^0}(inst)=(17-40)$ MeV 
compared to the leading order result $m_{K^0}(inst)=(85-120)$ MeV.
The large difference between the one and two-loop results is 
related to the fact that at moderate density perturbation theory
predicts that the average instanton size is not small. Using
equns.~(\ref{G_I},\ref{E_I}) we get
\be
\bar{\rho} = \frac{\Gamma(13/2)}{\Gamma(6)}
  \left(\sqrt{3}\mu\right)^{-1}
  \simeq 1.4\mu^{-1}.
\ee
For $\mu=500$ MeV we find $\bar{\rho}=0.55$ fm, which 
is bigger than the standard estimate for the typical instanton
size at zero density $\bar{\rho}_0=(0.3-0.4)$ fm. The 
situation is somewhat improved for the two-loop size
distribution which gives $\bar{\rho}=0.45$ fm. The 
main difference between the one and two-loop size distributions
is that at two-loop order the pre-exponent is evaluated at 
a scale given by the inverse instanton size $\rho^{-1}$ 
rather than at the external scale $\mu$. This difference 
is formally of higher order, but at moderate density it 
provides a significant suppression of large instantons. 
 
 We can also study this problem in a different way. It is 
clear that at small density there has to be some non-perturbative
effect that eliminates the contribution of large-size instantons.
We can simulate this effect in terms of a non-perturbative 
screening factor $\exp(-\rho^2m_{scr}^2)$ in the instanton
size distribution. If the screening mass is adjusted in such
a way that the average instanton size at $\mu=500$ MeV is 
equal to the phenomenological value at $\mu=0$, $\bar\rho=0.35$ 
fm, then we find $m_{K^0}(inst)=(7-12)$ MeV. 

 The sensitivity of the instanton contribution to the kaon
mass to large-size instantons translates into a large 
uncertainty regarding the behavior of kaons at finite 
density. If large instantons with $\rho\simeq 0.5$ fm
play a role then the kaon mass is dominated 
by the instanton contribution and kaon condensation 
is unlikely. If the typical instanton size satisfies
$\rho<0.35$ fm then the instanton contribution to the 
kaon mass is at most comparable to the perturbative
contribution and kaon condensation is likely.

%%%%%%%%%%%%%%%%%%%%%%%%%%%%%%%%%%%%%%%%%%%%%%%%%%%%%%%%%%%%%%%%%%%%%%%%%%%
\section{Finite instanton size effects}
\label{sec_size}
%%%%%%%%%%%%%%%%%%%%%%%%%%%%%%%%%%%%%%%%%%%%%%%%%%%%%%%%%%%%%%%%%%%%%%%%%

  In the previous section we computed the instanton contribution
to the vacuum energy using the effective instanton induced 
interaction equ.~(\ref{l_nf3}). This effective interaction is
derived under the assumption that the relevant momenta are smaller
than the inverse instanton size, $p,\mu\ll\rho^{-1}$. In our case
the relevant momenta are on the order of the Fermi momentum 
$|\vec{p}|\sim p_F \sim \mu$ and the typical instanton size
is given by $\rho\sim\mu^{-1}$, so $p\rho\sim 1$. This implies
that finite size effects are potentially important. 

  Instanton finite size effects can be determined from the 
fermion zero mode solution at finite baryon density. This 
solution was found in \cite{Abrikosov:rh} and studied in
\cite{Schafer:1998up,Rapp:1999qa,Carter:1999mt}. Using the 
exact zero mode solution amounts to the replacement
\be
 \bar{\psi}_{f\alpha i} \to \frac{1}{2\pi}
            \bar{\psi}_{f\alpha j} {\cal F}_i^j,\hspace{1cm}
 {\psi}^{f\alpha i} \to \frac{1}{2\pi}
          {\cal F^\dagger}^i_j {\psi}^{f\alpha j}
\ee
in the effective interaction equ.~(\ref{l_nf3}). Here, the instanton
form factors ${\cal F,\,F^\dagger}$ are given by 
\cite{Rapp:1999qa,Carter:1999mt}
\be
 {\cal F}_i^j = [(\tilde{p}\cdot\sigma^-)
                          (\varphi\cdot\sigma^+)]_i^j , \hspace{1cm}
 {\cal F^\dagger}^i_j = [ (\varphi^*\cdot\sigma^-)
                          (\tilde{p}\cdot\sigma^+)]^i_j 
\ee
with $\sigma^\pm_\mu=(\pm i\vec{\sigma},1)$, $\tilde{p}=
(\vec{p},p_4+i\mu)$, $\varphi_\alpha=\varphi_\alpha(p,\mu)$ and 
$\varphi^*_\alpha=[\varphi_\alpha(p,-\mu)]^*$. The instanton
form factors are determined by the Fourier transform 
$\varphi_\alpha(p,\mu)$ of the instanton zero mode wave function.
For completeness, we provide the result for $\varphi_\alpha
(p,\mu)$ in appendix A. 

  We first study how the instanton form factor modifies
the loop integral which contains the mass insertion, 
see equ.~(\ref{l_nf2}). In this case we need the quark-anti-quark 
form factor
\be
G_1(p) = \frac{1}{2(2\pi)^2}\delta^i_l\delta^j_k
  {\cal F}_i^k(p) {\cal F^\dagger}_j^l(p) .
\ee
We find $G_1(p)=(p+i\mu)^2\varphi_\alpha(p,\mu)\varphi_\alpha
(p,\mu)/(2\pi)^2$. We can now perform the loop integral with the 
mass insertion. We find
\be
 2N_c\int\frac{d^4p}{(2\pi)^4} \frac{mG_1(p)}{(p+i\mu)^2} = 
  \frac{N_c m}{(2\pi\rho)^2}\, ,
\ee
where we have used the normalization condition equ.~(\ref{norm})
for the Fourier transform of the zero mode wave function. We 
observe that the instanton form factor does not modify 
equ.~(\ref{l_nf2}).

\begin{figure}
\begin{center}
\leavevmode
\vspace*{-1cm}
\epsfxsize=12cm
\epsffile{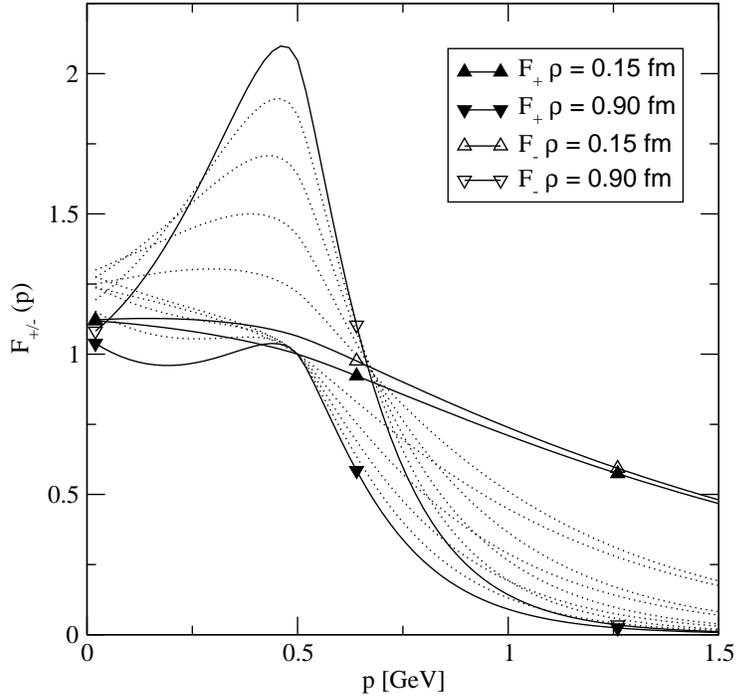}
\end{center}  
\caption{\label{fig_fpm}
Instanton form factors $F_\pm$ for quark-quark scattering. 
The form factors are shown as a function of $p=|\vec{p}|$
for $p_4=0$. The Fermi momentum was chosen to be $p_F=0.5$
GeV. The different curves correspond to instanton sizes 
in the range $\rho=(0.15-0.90)$ fm. }
\end{figure}

\begin{figure}
\begin{center}
\leavevmode
\vspace*{-1cm}
\epsfxsize=12cm
\epsffile{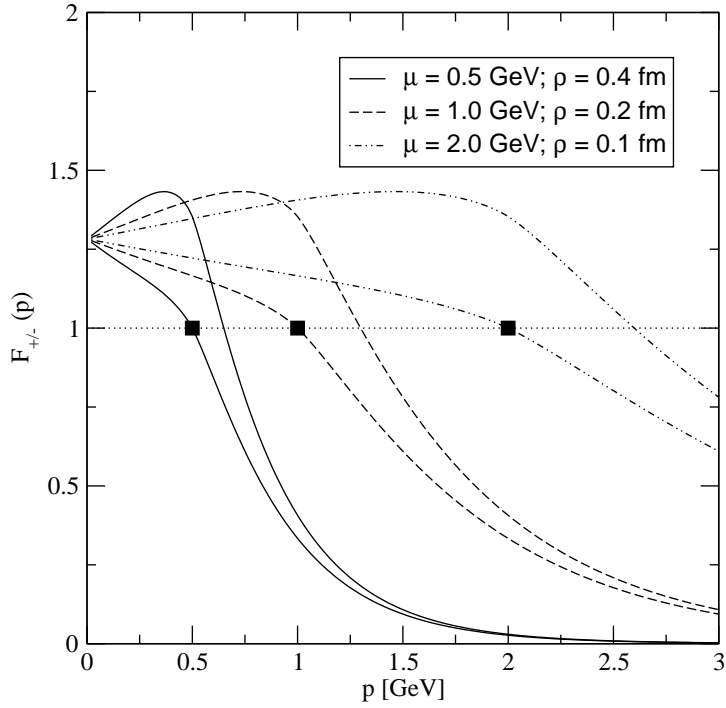}
\end{center}  
\caption{\label{fig_fpm_2}
Same as Fig.~\ref{fig_fpm} for different Fermi momenta $p_F
=0.5$, 1.0, 2.0 GeV. The upper and lower curves show $F_-$
and $F_+$, respectively. The solid square shows the value
of $F_+$ on the Fermi surface. The instanton size was fixed at $\rho
p_F=1$, corresponding to $\rho=0.4$, 0.2, 0.1 fm. }
\end{figure}

 The next step is the integration over the quark-quark 
propagators, equ.~(\ref{E_I}). The diquark loop involves
the form factors 
\bea
\label{def_f1}
 F_1(p) &=& \frac{1}{2(2\pi)^2}\epsilon^{ik}
  {\cal F}_i^j(p){\cal F}_k^l(-p)
   \epsilon_{jl}  \\
\label{def_f2}
 F_2(p) &=& \frac{1}{2(2\pi)^2}\epsilon^{ik}
  {\cal F}_i^j(p){\cal F}_k^l(-p)
   \epsilon_{jm} [(\sigma_0)(\vec{\sigma}\cdot\hat{p})]^m_l.
\eea
These two structures arise from the contraction of the instanton 
vertex with the diquark propagator
\be
\label{s12}
 S_{12}(p) = \frac{1}{2}(1+\vec{\alpha}\cdot\hat{p})C\gamma_5
 \frac{\Delta(|\vec{p}|)}{p_4^2+(|\vec{p}|-\mu)^2+\Delta^2}.
\ee
The two form factors $F_{1,2}(p)$ are given by
\bea
\label{f1}
 F_1(p) &=& \frac{-1}{(2\pi)^2}\Big\{ (p^2+\mu^2) 
                   \varphi(p,\mu)\cdot\varphi(-p,\mu)
  + (p+i\mu)\cdot\varphi(p,\mu) 
                   (p-i\mu)\cdot\varphi(-p,\mu) \nonumber \\
 & & \mbox{}\hspace{1.2cm} 
  - (p+i\mu)\cdot\varphi(-p,\mu) 
                   (p-i\mu)\cdot\varphi(p,\mu) \Big\}, \\
\label{f2}
 F_2(p) &=& \frac{i}{(2\pi)^2} \Big\{
    (p+i\mu)\cdot\varphi(p,\mu)\hspace{0.25cm}
  \left[ (p_4-i\mu) \hat{p}\cdot\vec{\phi}(-p,\mu)
        - p \varphi_4(-p,\mu) \right] \nonumber \\
        & &\hspace{0.7cm}\mbox{} 
        -(p-i\mu)\cdot\varphi(-p,\mu)
  \left[ (p_4+i\mu) \hat{p}\cdot\vec{\phi}(p,\mu)
        - p \varphi_4(p,\mu) \right]\Big\} .
\eea
The functions $F_{\pm}(p)=F_1(p)\pm F_2(p)$ are shown in 
Figs.~\ref{fig_fpm},\ref{fig_fpm_2}. We observe that the form factors 
are centered around the Fermi surface, $p_4=0,\,|\vec{p}|=\mu$. 
This is an important observation. Instantons produce 
$(\bar{\psi}_R\psi_L)$-particle-hole pairs near the Fermi 
surface, and the chirality violating pair creation amplitude is 
therefore not suppressed by Pauli-blocking \cite{Rapp:1999qa}.
In Fig.~\ref{fig_fpm} we study the dependence of $F_\pm(p)$ on 
$\rho$ at fixed baryon chemical potential $\mu$. We find that on 
the Fermi surface $F_+=1$ for all $\rho$ whereas $F_-\to 1$ 
only as $\rho\to 0$. In practice we are interested in the limit 
$\mu\to\infty$ with $\rho\sim\mu^{-1}$. This limit is studied 
in Fig.~\ref{fig_fpm_2}. Again we observe that $F_+(p\!=\!\mu,
p_4\!=\!0)=1$ for all $\mu$. 

  In the weak coupling limit the contraction of the diquark 
propagator equ.~(\ref{s12}) with the instanton vertex only
involves the $F_+$ form factor. Expanding $F_+(p)$ around the 
Fermi surface and using  $F_+(\mu,0)=1$ we can show that 
up to corrections of order $O(g^2)$ the instanton form factor
does not modify the weak coupling result equ.~(\ref{A}) for $A$.
In the same fashion we can also show that up to corrections
of order $O(g^2)$ there is no anti-gap ($F_-$) contribution 
to $A$.

\begin{figure}
\begin{center}
\leavevmode
\vspace*{-1cm}
\epsfxsize=12cm
\epsffile{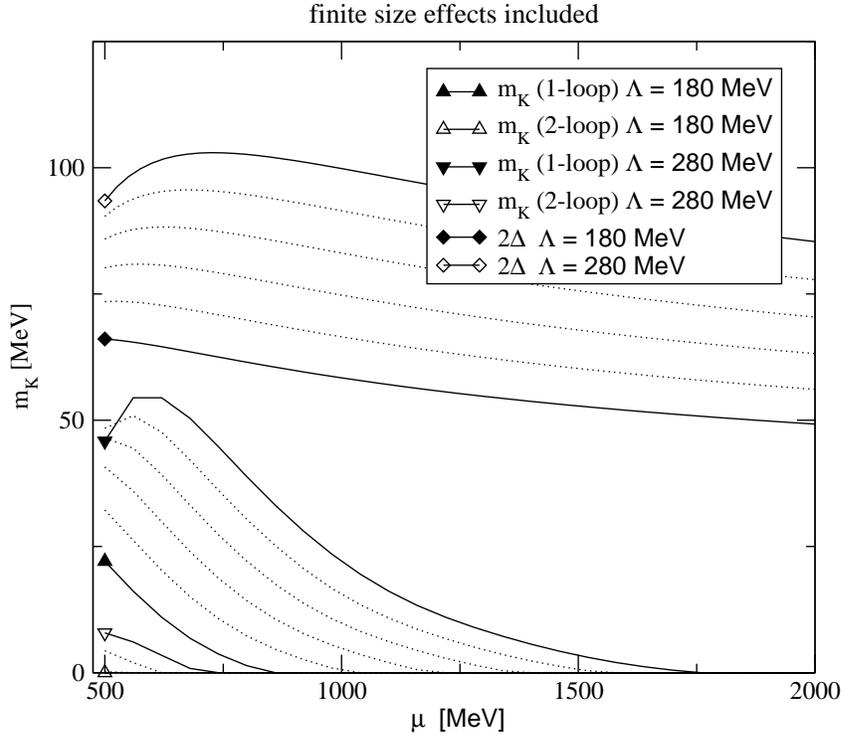}
\end{center}  
\caption{\label{fig_mgb_3}
Same as Fig.~\ref{fig_mgb_2} but with instanton finite 
size effects (instanton form factors) included.}
\end{figure}

 We have also studied the effect of the instanton form factor 
at moderate density. The integral over the diquark propagator 
is given by 
\be 
\label{qq_ff}
 \Phi = \int\frac{d^4p}{(2\pi)^4} \frac{F_+(p)\Delta(|\vec{p}|)}
   {p_4^2+(|\vec{p}|-\mu)^2+\Delta^2}.
\ee
For $F_+(p)=1$ we recover the perturbative 
result equ.~(\ref{qq_cond}). Results for the kaon mass with 
the effects of the instanton form factor included are shown 
in Fig.~\ref{fig_mgb_3}. At a baryon chemical potential 
$\mu=500$ MeV the instanton form factor leads to a modest 
reduction of the instanton contribution to the kaon mass 
on the order of $40\%$. This reduction is due to the 
fact that the instanton form factor suppresses the contribution
from large instantons. This effect, however, does not 
lead to a significant reduction of the average instanton
size. In particular, the discrepancy between the one 
and two-loop results remains large.

%%%%%%%%%%%%%%%%%%%%%%%%%%%%%%%%%%%%%%%%%%%%%%%%%%%%%%%%%%%%%%%%%%%%%%%
\section{QCD with two colors}
\label{sec_nc2}
%%%%%%%%%%%%%%%%%%%%%%%%%%%%%%%%%%%%%%%%%%%%%%%%%%%%%%%%%%%%%%%%%%%%%%%%%

  Given the large uncertainty in the instanton contribution 
to the Goldstone boson masses at moderate baryon density it would
clearly be useful if the result could be checked on the lattice. 
Because of the sign problem direct studies of $N_c=3$ QCD are 
still not feasible. On the other hand, QCD with $N_c=2$
colors and $N_f=2$ flavors does not suffer from a sign problem at 
non-zero baryon density and has been studied successfully on the 
lattice \cite{Dagotto:1986gw,Hands:1999md,Kogut:2001na}.

 In QCD with two colors we also expect the formation of a 
diquark condensate at large baryon density \cite{Rapp:1998zu}. 
For $N_f=2$ flavors the diquark condensate breaks the $U(1)_B$
of baryon number and the anomalous $U(1)_A$ symmetry, but 
not the $SU(2)_L\times SU(2)_R$ chiral symmetry. The $U(1)_B$
Goldstone boson is exactly massless, but the $U(1)_A$ 
Goldstone boson acquires a mass from instantons. The 
corresponding contribution to the vacuum energy is shown 
in Fig.~\ref{fig_mvac}f. Except for the mass insertion, 
the diagram is identical to the $O(m)$ contribution to
the vacuum energy in the CFL phase of $N_c=N_f=3$ QCD.

 The effective lagrangian for the pseudoscalar Goldstone boson 
is 
\be
 {\cal L} = \frac{f^2}{2}\left[ (\partial_0\phi)^2-v^2(\partial_i\phi)^2
 \right] -  V(\phi).
\ee
We can think of $\phi$ as the relative phase between the 
left and right handed diquark condensate. Since baryon 
number is broken, the field $\phi$ has the quantum numbers
of both pseudoscalar mesons and diquarks. This means that 
the mass of the $\phi$-field governs the asymptotic behavior
of both the $\eta'$ and pseudoscalar diquark correlation
functions. The decay constant $f$ and Goldstone boson velocity 
$v$ can be determined in perturbation theory. The result
is identical to the one in QCD with $N_c=3$ colors and $N_f=2$
flavors \cite{Son:1999cm,Beane:2000ms}
\be
\label{f_nc2}
 f^2 = \left( \frac{\mu^2}{4\pi^2}\right),
 \hspace{1cm} v^2 = \frac{1}{3}.
\ee
The potential $V(\phi)$ is determined by instantons. The 
calculation of the instanton induced potential is completely
analogous to the case $N_c=3$, $N_f=2$ \cite{Son:2001jm}, so 
we can be brief here. We find $V(\phi)=A_2\cos(\phi-\theta)$ 
where $\theta$ is the QCD theta angle and
\be 
\label{A2}
 A_2 = C_N 8\pi^4 \left[\frac{4\pi}{g}\Delta
     \left(\frac{\mu^2}{2\pi^2}\right)\right]^2
  \left(\frac{8\pi^2}{g^2}\right)^{4}
  \left(\frac{\Lambda}{\mu}\right)^{8}\Lambda^{-2}.
\ee
In $N_c=2$ QCD the gap $\Delta$ is given by
\be 
\label{gap_nc2}
 \Delta = 512\pi^4 b_0'\mu g^{-5}
 \exp\left(-\frac{2\pi^2}{g}\right).
\ee 
Using these results we can determine the mass of the 
pseudoscalar Goldstone boson. We find
\be
\label{m_nc2}
 m_\phi^2 = \frac{A_2}{f^2}.
\ee
This result looks like the Witten-Veneziano
relation \cite{Witten:1979vv,Veneziano:1979ec}
\be 
\label{WV}
 m_{\eta'}^2 = \frac{2N_f\chi_{top}}{f_{\eta'}^2},
\ee
where $A_2$ plays the role of the topological 
susceptibility\footnote{The extra factor $2N_f$ is related 
to our normalization of $f$. If we define $f_{\eta'}$ in such 
a way that the canonical axial current $A_0=\bar{\psi}\gamma_0
\gamma_5\frac{\tau^0}{2}\psi$ is represented in the 
effective theory by $A_0=f_{\eta'}\partial_0\eta'+\ldots$ 
then we find $m_{\eta'}^2f_{\eta'}^2=4A_2$.}. 
We note, however, that the topological susceptibility vanishes
in the chiral limit, $\chi_{top}=O(m^2)$. Indeed, $A_2$ is not the 
topological susceptibility but the density of instantons,
$A_2=(N/V)$. This is consistent with the idea that in a dilute 
system of instantons local fluctuations of the topological charge 
are Poissonian. This means that $\chi_{top}(V)=\langle Q^2\rangle/V
= (N/V)$ for any 4-volume $V$ which is large compared to the average 
volume per instanton and small compared to the screening volume
\cite{Shuryak:1994rr,Schafer:1996wv}, 
\be
\label{debye}
(N/V)^{-1}\ll V \ll r_D^4,
\ee
where $r_D\sim m_{\eta'}^{-1}$ is the screening length.
Because of the large mass of the $\eta'$ this window 
does not exist in QCD at zero density. In QCD at 
asymptotically high baryon density the decay constant 
scales as the baryon chemical potential, $f\sim\mu$,
the $\eta'$ mass is much smaller than $(N/V)^{1/4}$,
and the window described by equ.~(\ref{debye}) opens 
up \cite{Son:2001jm}.

 Mass relations of the Witten-Veneziano type can also 
be derived in QCD with $N_c=3$ colors. The case of
$N_f=2$ flavors is exactly analogous to the $N_c=N_f=2$
case except that the numerical value of the finite 
volume susceptibility $\chi_{top}(V)$ is different
\cite{Son:2001jm}. In the $N_f=3$ CFL phase we 
also find that the $\eta'$-mass in the chiral limit
satisfies a finite-volume Witten-Veneziano relation,
$m_{\eta'}^2 f_{\eta'}^2=6\chi_{top}(V)$. In this case, 
however, we have $\chi_{top}(V)=2(N/V)$ because instantons
occur in pairs. 

  We can also study how the relation between the $\eta'$ 
mass and the topological susceptibility is modified if 
quark masses are included. In the regime in which the 
$\eta'$ mass is dominated by the linear mass term, 
equ.~(\ref{m_eta}), we find $m_{\eta'}^2f_{\eta'}^2
=(4/9)\chi_{top}(V)$. In this regime the topological 
charge is only partially screened, $\chi_{top}(V\to
\infty)=\frac{1}{9}\chi_{top}(V)$. If the quark mass
is even larger and the quadratic mass term dominates
the relation between the $\eta'$ mass and the 
topological susceptibility is lost.
 
\begin{figure}
\begin{center}
\leavevmode
\vspace*{-1cm}
\epsfxsize=12cm
\epsffile{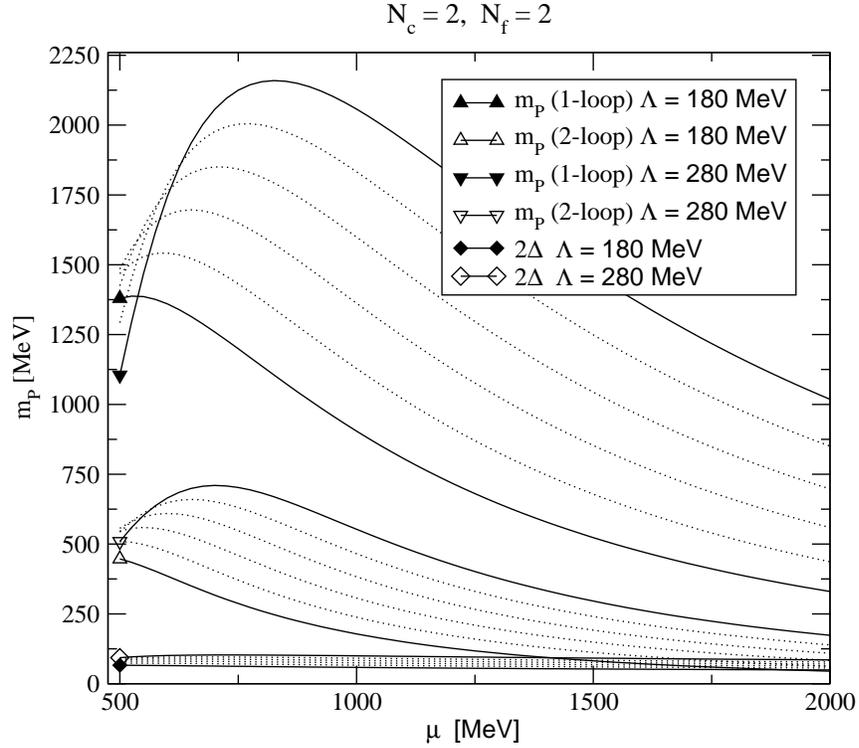}
\end{center}  
\caption{\label{fig_nc2}
Pseudoscalar Goldstone boson mass in $N_c=2$ QCD. The two
sets of curves correspond to the one and two-loop instanton
size distribution and different values of the scale
parameter, see Fig.~\ref{fig_mgb_1}. For comparison, we also 
show the energy gap $2\Delta$. }
\end{figure}

\begin{figure}
\begin{center}
\leavevmode
\vspace*{-1cm}
\epsfxsize=12cm
\epsffile{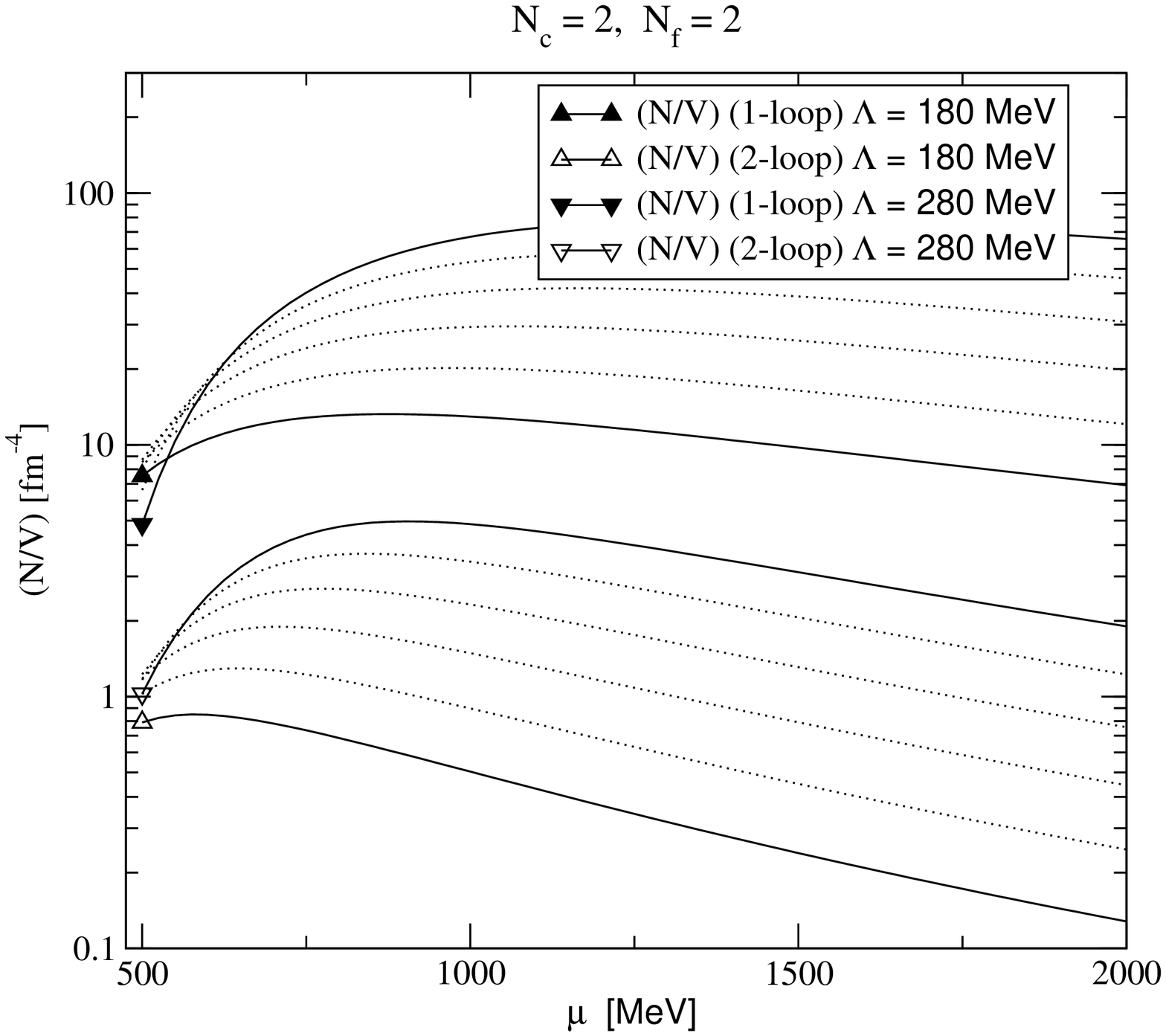}
\end{center}  
\caption{\label{fig_nc2_2}
Instanton density in $N_c=2$ QCD. The two
sets of curves correspond to the one and two-loop instanton
size distribution and different values of the scale parameter
in the range $\Lambda=(180-280)$ MeV. }
\end{figure}

 In Figs.~\ref{fig_nc2},\ref{fig_nc2_2} we show numerical
results for the $\eta'$ mass and the instanton density in
QCD with $N_c=N_f=2$. We again find that at moderate 
densities, $\mu\simeq 500$ MeV, the result is dominated
by relatively large instantons $\bar{\rho}\simeq 0.5$ fm.
The one-loop results for both the $\eta'$ mass and the 
instanton density are surprisingly large. Both are bigger 
than the phenomenological values $m_{\eta'}=945$ MeV and
$(N/V)\simeq 1\,{\rm fm}^{-4}$ in $N_c=3$ QCD at zero 
baryon density. This is incompatible with the idea
that $m_{\eta'}$ and $(N/V)$ decrease as a function
of density, and that the dependence on $N_c$ is weak. 
Lattice simulations of $N_c=2$ also point to a very
light pseudoscalar diquark \cite{Hands:1999md,Kogut:2001na}.
We should note, however, that these simulations were
performed with more than two continuum quark flavors.

%%%%%%%%%%%%%%%%%%%%%%%%%%%%%%%%%%%%%%%%%%%%%%%%%%%%%%%%%%%%%%%%%%%%%%%%%
\section{Summary}
\label{sec_sum}
%%%%%%%%%%%%%%%%%%%%%%%%%%%%%%%%%%%%%%%%%%%%%%%%%%%%%%%%%%%%%%%%%%%%%%%%%

  We have computed the instanton contribution to the masses of 
Goldstone bosons in the CFL phase of QCD with three quark flavors. 
Our main result is given in equns.~(\ref{A}) and (\ref{mgb}). At 
very large baryon density this result is expected to be exact.

  At densities $\rho\simeq (5-10)\rho_0$ that are of interest in 
connection with the physics of neutron stars the perturbative
result is very sensitive to the contribution of large instantons. 
For $\mu=500$ MeV leading order perturbation theory predicts
$\bar{\rho}=0.55$ fm and $m_{K}(inst)=(85-120)$ MeV. If this 
result is reliable, then instantons would almost certainly prevent 
kaon condensation in the CFL phase at densities $\rho<10\rho_0$. 
The situation is different if the two-loop instanton size 
distribution is used. In this case we find $\bar{\rho}=0.45$
fm and $m_{K}(inst)=(17-40)$ MeV. The result is further reduced
if a phenomenological screening mass and the instanton form factor 
is taken into account. In this case we find $m_{K}(inst)=(7-12)$ 
MeV. 

 We suggest that these uncertainties can be addressed 
using lattice calculations of the instanton density and 
the pseudoscalar diquark mass in $N_c=2$ QCD. The result
of the leading order calculation, equ.~(\ref{WV})
suggests that the pseudoscalar diquark mass is much bigger 
than $2\Delta$. This prediction is only weakly dependent
on the value of the scale parameter and the magnitude of
the gap $\Delta$. We also emphasize that the pseudoscalar
diquark mass is related to the topological susceptibility
in a finite volume. 

Acknowledgments: We would like to thank M.~Alford, P.~Bedaque, 
G.~Carter, C.~Manuel, D.~Son, and M.~Stephanov for useful discussions. 
This work was supported in part by US DOE grant DE-FG-88ER40388. 

\newpage
\appendix
\section{Fourier Transform of Fermion Zero Modes}
\label{sec_app}

 We repeat the results for the Fourier transform 
of the fermion zero mode wave function \cite{Rapp:1999qa,Carter:1999mt}.
We follow the notation of \cite{Carter:1999mt}. The Fourier
transform is given by
\be
\psi_{L,R}(p,\mu)^{\alpha i} = \varphi_\nu(p,\mu)
        \left(\sigma^{\pm}_{\nu}\right)^i_j
        \epsilon^{jk} U^{\alpha}_k \,.
\ee
The temporal and spatial components of $\varphi_\nu$ are 
given by
\bea
\varphi_4(p_4,p,\mu) &=&  \frac{\pi\rho^2}{4p} \Bigg\{
  \left(p-\mu-ip_4\right)
  \left[\left(2p_4+i\mu\right)f_{1-} 
      + i\left(p-\mu-ip_4\right)f_{2-}\right]
  \nonumber\\
  & & \mbox{}\hspace{0.5cm} 
      + \left(p+\mu+ip_4\right)
        \left[\left(2p_4+i\mu\right)f_{1+} -
         i\left(p+\mu+ip_4\right)f_{2+}\right]\Bigg\}\,, \\
\vec{\varphi}(p_4,p,\mu) &=& \frac{\pi\rho^2 \hat{p}_i}{4p} \Bigg\{
  (2p-\mu)\left(p-\mu-ip_4\right)f_{1-}
  +(2p+\mu)\left(p+\mu+ip_4\right)f_{1+} 
  \nonumber\\
  & &  \mbox{}\hspace{0.5cm} + \left( 2(p-\mu)
                   \left(p-\mu-ip_4\right) -
  \frac{1}{p}\left(\mu+ip_4\right)
         \left[p_4^2+(p-\mu)^2\right] \right) f_{2-} 
  \nonumber\\
  & &   \mbox{}\hspace{0.5cm}+
  \left( 2(p+\mu)\left(p+\mu+ip_4\right) +
  \frac{1}{p}\left(\mu+ip_4\right)
    \left[p_4^2+(p+\mu)^2\right] \right) f_{2+}\Bigg\}\, 
\eea
where $p=|\vec{p}|$, $\hat{p}=\vec{p}/p$ and we have introduced
the functions
\bea
f_{1\pm} &=& \frac{1}{z_\pm}
 \left[I_1(z_{\pm})K_0(z_{\pm}) - I_0(z_{\pm}) K_1(z_{\pm})\right],\\
f_{2\pm} &=& \frac{1}{z_\pm^2}
  I_1(z_{\pm})K_1(z_{\pm}).
\eea
The argument of $f_{1,2\pm}$ is given by $z_{\pm} = \frac{1}{2}
\rho\sqrt{p_4^2+(p\pm\mu)^2}$. The Fourier transform $\varphi_\nu
(p,\mu)$ satisfies the normalization condition
\be
\label{norm}
 \int\frac{d^4p}{(2\pi)^4} \varphi_\nu(p,\mu)\varphi_\nu(p,\mu)
=\int\frac{d^4p}{(2\pi)^4} [\varphi_\nu(p,-\mu)]^*\varphi_\nu(p,\mu)
=\frac{1}{2\rho^2}\, .
\ee

\newpage 

\newpage 

\newpage

\newpage 

\newpage

\newpage 

\newpage

\begin{thebibliography}{20}

\bibitem{Alford:1999mk}
M.~Alford, K.~Rajagopal and F.~Wilczek,
%``Color-flavor locking and chiral symmetry breaking in high density {QCD},''
Nucl.\ Phys.\  {\bf B537}, 443 (1999)
[hep-ph/9804403].
%%CITATION = HEP-PH 9804403;%%

\bibitem{Schafer:1999fe}
T.~Sch{\"a}fer,
%``Patterns of symmetry breaking in QCD at high baryon density,''
Nucl.\ Phys.\ {\bf B575}, 269 (2000)
[hep-ph/9909574].
%%CITATION = HEP-PH 9909574;%%
     
\bibitem{Evans:2000at}
N.~Evans, J.~Hormuzdiar, S.~D.~Hsu and M.~Schwetz,
%``On the QCD ground state at high density,''
Nucl.\ Phys.\ B {\bf 581}, 391 (2000)
[hep-ph/9910313].
%%CITATION = HEP-PH 9910313;%%

\bibitem{Schafer:1999ef}
T.~Sch{\"a}fer and F.~Wilczek,
%``Continuity of quark and hadron matter,''
Phys.\ Rev.\ Lett.\  {\bf 82}, 3956 (1999)
[hep-ph/9811473].
%%CITATION = HEP-PH 9811473;%%

\bibitem{Son:1999cm}
%``Inverse Meson Mass Ordering in CFL Phase of QCD''
D.~T.~Son and M.~Stephanov, 
Phys.\ Rev.\ {\bf D61}, 074012 (2000) 
[hep-ph/9910491], 
erratum: hep-ph/0004095.
%%CITATION = HEP-PH 9910491;%%

\bibitem{Rho:2000xf}
M.~Rho, A.~Wirzba, and I.~Zahed,
%``Generalized pions in dense QCD,''
Phys.\ Lett.\ {\bf B473}, 126 (2000)
[hep-ph/9910550].
%%CITATION = HEP-PH 9910550;%%

\bibitem{Hong:2000ei}
D.~K.~Hong, T.~Lee, and D.~Min, 
%``Meson mass at large baryon chemical potential in dense QCD,''
Phys.\ Lett.\ {\bf B477}, 137 (2000)
[hep-ph/9912531].
%%CITATION = HEP-PH 9912531;%%

\bibitem{Manuel:2000wm}
C.~Manuel and M.~H.~Tytgat,
%``Masses of the Goldstone modes in the CFL phase of QCD at finite  density,''
Phys.\ Lett.\ {\bf B479}, 190 (2000)
[hep-ph/0001095].
%%CITATION = HEP-PH 0001095;%%

\bibitem{Rho:2000ww}
M.~Rho, E.~Shuryak, A.~Wirzba and I.~Zahed,
%``Generalized mesons in dense QCD,''
Nucl.\ Phys.\ A {\bf 676}, 273 (2000)
[hep-ph/0001104].
%%CITATION = HEP-PH 0001104;%%

\bibitem{Beane:2000ms}
S.~R.~Beane, P.~F.~Bedaque, and M.~J.~Savage,
%``Meson masses in high density QCD,''
Phys.\ Lett.\ {\bf B483}, 131 (2000)
[hep-ph/0002209].
%%CITATION = HEP-PH 0002209;%%

\bibitem{Hong:2000ng}
D.~K.~Hong,
%``Radiative mass in QCD at high density,''
Phys.\ Rev.\ D {\bf 62}, 091501 (2000)
[hep-ph/0006105].
%%CITATION = HEP-PH 0006105;%%

\bibitem{Schafer:2001za}
T.~Sch\"afer,
%``Mass Terms in Effective Theories of High Density Quark Matter,''
Phys.\ Rev.\ {\bf D}, in press, 
hep-ph/0109052.
%%CITATION = HEP-PH 0109052;%%

\bibitem{Alford:1999pa}
M.~Alford, J.~Berges and K.~Rajagopal,
%``Unlocking color and flavor in superconducting strange quark matter,''
Nucl.\ Phys.\  {\bf B558}, 219 (1999)
[hep-ph/9903502].
%%CITATION = HEP-PH 9903502;%%
 
\bibitem{Schafer:1999pb}
T.~Sch{\"a}fer and F.~Wilczek,
%``Quark description of hadronic phases,''
Phys.\ Rev.\  {\bf D60}, 074014 (1999)
[hep-ph/9903503].
%%CITATION = HEP-PH 9903503;%% 

\bibitem{Alford:2001ze}
M.~Alford, J.~Bowers and K.~Rajagopal,
%``Crystalline color superconductivity,''
Phys.\ Rev.\ D {\bf 63}, 074016 (2001)
[hep-ph/0008208].
%%CITATION = HEP-PH 0008208;%%

\bibitem{Schafer:2000ew}
T.~Sch{\"a}fer,
%``Kaon condensation in high density quark matter,''
Phys.\ Rev.\ Lett.\ {\bf 85}, 5531 (2000)
[nucl-th/0007021].
%%CITATION = NUCL-TH 0007021;%%

\bibitem{Bedaque:2001je}
P.~F.~Bedaque and T.~Sch{\"a}fer,
%``High density quark matter under stress,''
Nucl.\ Phys.\ {\bf A697}, 802 (2002)
[hep-ph/0105150].
%%CITATION = HEP-PH 0105150;%%

\bibitem{Alford:2001zr}
M.~Alford, K.~Rajagopal, S.~Reddy and F.~Wilczek,
%``The minimal CFL nuclear matter interface''
Phys.\ Rev.\ D {\bf 64}, 074017 (2001)
[hep-ph/0105009].
%%CITATION = HEP-PH 0105009;%%

\bibitem{Kaplan:2001qk}
D.~B.~Kaplan and S.~Reddy,
%``Novel phases and transitions in quark matter,''
hep-ph/0107265.
%%CITATION = HEP-PH 0107265;%%

\bibitem{Rapp:1999qa}
R.~Rapp, T.~Sch\"afer, E.~V.~Shuryak, and M.~Velkovsky,
%``High Density QCD and Instantons''
Annals Phys.\ {\bf 280}, 35 (2000),
[hep-ph/9904353].
%%CITATION = HEP-PH 9904353;%%

\bibitem{Son:2001jm}
D.~T.~Son, M.~A.~Stephanov and A.~R.~Zhitnitsky,
%``Instanton interactions in dense matter QCD''
Phys.\ Lett.\ B {\bf 510}, 167 (2001)
[hep-ph/0103099].
%%CITATION = HEP-PH 0103099;%%

\bibitem{Casalbuoni:1999wu}
R.~Casalbuoni and D.~Gatto,
%``Effective theory for color-flavor locking in high density QCD,''
Phys.\ Lett.\ {\bf B464}, 111 (1999)
[hep-ph/9908227].
%%CITATION = HEP-PH 9908227;%%

% SYMMETRY BREAKING THROUGH ABJ ANOMALIES
\bibitem{'tHooft:up}
G.~'t Hooft,
Phys.\ Rev.\ Lett.\ {\bf 37}, 8 (1976).
%%CITATION = PRLTA,37,8;%%

% INSTANTON DENSITY IN A THEORY WITH MASSLESS QUARKS
\bibitem{Shifman:uw}    
M.~A.~Shifman, A.~I.~Vainshtein and V.~I.~Zakharov,
Nucl.\ Phys.\ B {\bf 163}, 46 (1980).
%%CITATION = NUPHA,B163,46;%%

% INSTANTONS IN QCD
\bibitem{Schafer:1996wv} 
T.~Sch{\"a}fer and E.~V.~Shuryak, 
Rev.\ Mod.\ Phys.\ {\bf 70}, 323 (1998)
[hep-ph/9610451].
%%CITATION = HEP-PH 9610451;%%

\bibitem{Son:1999uk}
D.~T.~Son,
%``Superconductivity by long-range color magnetic interaction in  high-density quark matter,''
Phys.\ Rev.\ D {\bf 59}, 094019 (1999)
[hep-ph/9812287].
%%CITATION = HEP-PH 9812287;%%

\bibitem{Schafer:1999jg}
T.~Sch{\"a}fer and F.~Wilczek,
%``Superconductivity from perturbative one-gluon exchange in high density  quark matter,''
Phys.\ Rev.\  {\bf D60}, 114033 (1999)
[hep-ph/9906512].
%%CITATION = HEP-PH 9906512;%%
  
\bibitem{Hong:2000fh}
D.~K.~Hong, V.~A.~Miransky, I.~A.~Shovkovy and L.~C.~Wijewardhana,
%``Schwinger-Dyson approach to color superconductivity in dense QCD,''
Phys.\ Rev.\  {\bf D61}, 056001 (2000)
[hep-ph/9906478].
%%CITATION = HEP-PH 9906478;%%

%\cite{Pisarski:2000tv}
\bibitem{Pisarski:2000tv}
R.~D.~Pisarski and D.~H.~Rischke,
%``Color superconductivity in weak coupling,''
Phys.\ Rev.\  {\bf D61}, 074017 (2000)
[nucl-th/9910056].
%%CITATION = NUCL-TH 9910056;%%

\bibitem{Brown:1999aq}
W.~E.~Brown, J.~T.~Liu and H.~Ren,
%``On the perturbative nature of color superconductivity,''
Phys.\ Rev. {\bf D61}, 114012 (2000)
[hep-ph/9908248].
%%CITATION = HEP-PH 9908248;%%

\bibitem{Wang:2001aq}
Q.~Wang and D.~H.~Rischke,
%``How the quark self-energy affects the color-superconducting gap,''
nucl-th/0110016.
%%CITATION = NUCL-TH 0110016;%%

\bibitem{Zarembo:2000pj}
K.~Zarembo,
%``Dispersion laws for Goldstone bosons in a color superconductor,''
Phys.\ Rev.\ D {\bf 62}, 054003 (2000)
[hep-ph/0002123].
%%CITATION = HEP-PH 0002123;%%

\bibitem{Miransky:2001bd}
V.~A.~Miransky, I.~A.~Shovkovy and L.~C.~Wijewardhana,
%``Bethe-Salpeter equation for diquarks in cfl phase of cold dense QCD,''
Phys.\ Rev.\ D {\bf 63}, 056005 (2001)
[hep-ph/0009173].
%%CITATION = HEP-PH 0009173;%%

\bibitem{Abrikosov:rh}
A.~A.~Abrikosov,
%``Instantons In Quark Plasma, Particle Propagator And Zero Fermion Modes. (In Russian),''
Yad.\ Fiz.\  {\bf 37}, 772 (1983).
%%CITATION = YAFIA,37,772;%%

\bibitem{Schafer:1998up}
T.~Sch{\"a}fer,
%``Instantons and the chiral phase transition at non-zero baryon density,''
Phys.\ Rev.\ D {\bf 57}, 3950 (1998)
[hep-ph/9708256].
%%CITATION = HEP-PH 9708256;%%

\bibitem{Carter:1999mt}
G.~W.~Carter and D.~I.~Diakonov,
%``Instanton induced interactions in finite density QCD''
Nucl.\ Phys. A {\bf 661}, 625 (1999),
[hep-ph/9908314].
%%CITATION = HEP-PH 9908314;%%
  
\bibitem{Rapp:1998zu}
R.~Rapp, T.~Sch{\"a}fer, E.~V.~Shuryak and M.~Velkovsky,
%``Diquark Bose condensates in high density matter and instantons,''
Phys.\ Rev.\ Lett.\  {\bf 81}, 53 (1998)
[hep-ph/9711396].
%%CITATION = HEP-PH 9711396;%% 

\bibitem{Dagotto:1986gw}
E.~Dagotto, F.~Karsch and A.~Moreo,
%``The Strong Coupling Limit Of SU(2) QCD At Finite Baryon Density,''
Phys.\ Lett.\ B {\bf 169}, 421 (1986).
%%CITATION = PHLTA,B169,421;%%

\bibitem{Hands:1999md}
S.~Hands, J.~B.~Kogut, M.~P.~Lombardo and S.~E.~Morrison,
%``Symmetries and spectrum of SU(2) lattice gauge theory at finite  chemical potential,''
Nucl.\ Phys.\ B {\bf 558}, 327 (1999)
[hep-lat/9902034].
%%CITATION = HEP-LAT 9902034;%%

\bibitem{Kogut:2001na}
J.~B.~Kogut, D.~K.~Sinclair, S.~J.~Hands and S.~E.~Morrison,
%``Two-colour QCD at non-zero quark-number density,''
Phys.\ Rev.\ D {\bf 64}, 094505 (2001)
[hep-lat/0105026].
%%CITATION = HEP-LAT 0105026;%%

\bibitem{Witten:1979vv}
E.~Witten,
%``Current Algebra Theorems For The U(1) 'Goldstone Boson',''
Nucl.\ Phys.\ B {\bf 156}, 269 (1979).
%%CITATION = NUPHA,B156,269;%%

\bibitem{Veneziano:1979ec}
G.~Veneziano,
%``U(1) Without Instantons,''
Nucl.\ Phys.\ B {\bf 159}, 213 (1979).
%%CITATION = NUPHA,B159,213;%%

\bibitem{Shuryak:1994rr}
E.~V.~Shuryak and J.~J.~Verbaarschot,
%``Screening of the topological charge in a correlated instanton vacuum,''
Phys.\ Rev.\ D {\bf 52}, 295 (1995)
[hep-lat/9409020].
%%CITATION = HEP-LAT 9409020;%%

\end{thebibliography}
\end{document}